\documentclass[11pt]{article}
\usepackage{graphicx}
\newlength{\vshift}
\input epsf.tex
\newlength{\hshift}

\usepackage{amssymb}
\usepackage{amsmath}
\usepackage{amsthm}
\usepackage{amsopn}
\usepackage{color}
\usepackage[font=small,skip=0pt]{caption}
\newcommand{\cmmnt}[1]{}
\begin{document}

\title{\vspace*{-2.0cm}
\vglue -0.3cm
\bf\Large
High Energy Neutrinos from Recent Blazar Flares}
\author{
Francis Halzen\thanks{email: \tt francis.halzen@icecube.wisc.edu}\, and Ali Kheirandish\thanks{email: \tt ali.kheirandish@icecube.wisc.edu}
\\ \\
{\normalsize \it Wisconsin IceCube Particle Astrophysics Center and Department of Physics,}\\
{\normalsize \it University of Wisconsin, Madison, WI 53706, USA} \\
} 
\date{}
\maketitle
\thispagestyle{empty}
\vspace*{-1.0cm}
\vskip 2em
\begin{abstract}
 The energy density of cosmic neutrinos measured by IceCube matches the one observed by Fermi in extragalactic photons that predominantly originate in blazars. This has inspired attempts to match Fermi sources with IceCube neutrinos. A spatial association combined with a coincidence in time with a flaring source may represent a smoking gun for the origin of the IceCube flux. In June 2015, the Fermi Large Area Telescope observed an intense flare from blazar 3C 279 that exceeded the steady flux of the source by a factor of forty for the duration of a day. We show that IceCube is likely to observe neutrinos, if indeed hadronic in origin, in data that are still blinded at this time. We also discuss other opportunities for coincident observations that include a recent flare from blazar 1ES 1959+650 that previously produced an intriguing coincidence with AMANDA observations. 
\end{abstract}
\newpage
\section{Introduction}

Since the discovery of cosmic neutrinos \cite{science paper}, IceCube has been accumulating observations of these events \cite{IC 3rd year paper,Chris, Diffuse ICRC}. The higher statistics data reinforce the observation that the flux is predominantly extragalactic and reveal a flux of neutrinos with a total energy density that matches the one observed by Fermi in extragalactic gamma rays. This has bolstered the speculation that blazars, which are responsible for the majority of Fermi photons, are the sources of cosmic neutrinos.

Blazars are a subclass of active Galactic nuclei (AGN) with collimated jets aligned with the line of sight of the observer. With gamma-ray bursts, they have been widely speculated to be the sources of the highest energy cosmic rays and of accompanying neutrinos and gamma rays of pionic origin. Recent studies with the Fermi Large Area Telescope (Fermi LAT) have shown that blazars are responsible for more than 85\% of the extragalactic isotropic gamma-ray background (IGB) \cite{Ajelo}. From that along with the fact that a gamma-ray flux from neutral pions, accompanying the flux of the charged pions responsible for the IceCube neutrinos, matches the Fermi flux, blazars emerge as a plausible source of cosmic neutrinos \cite{Keith and Markus}. Also, recent studies have argued for a correlation between cosmic neutrinos and blazar catalogs  \cite{resconi}. Because blazars are flaring sources coincident in time as well in direction, they provide a powerful opportunity to make the case for such a connection, possibly with a single observation. The recent association of the second highest energy neutrino event of 2 PeV with the blazar PKS B1424-418 provides an interesting hint in this context \cite{TANAMI}.
 
The blazar spectral energy distribution generically has two components with two peaks in the IR/X-ray and the MeV/TeV photon energy ranges. The two components can typically be described in {\it leptonic} and {\it hadronic} scenarios where the acceleration of electrons and protons, respectively, are the origin of the high-energy photons. In the leptonic scenario, synchrotron radiation by electrons is responsible for the first peak, and Inverse Compton scattering on electrons produces the second \cite{Leptonics}. In hadronic models \cite{Hadronics}, both protons and electrons are accelerated. Synchrotron radiation still produces the low-energy peak in the spectrum, while the high-energy MeV/TeV photons are the decay products of pions produced in $pp$ or $p\gamma$ interactions in the jet. In one model, the protons interact with the synchrotron photons, for instance. The charged pions inevitably produced with neutral pions will be the parents of neutrinos that provide incontrovertible evidence of cosmic-ray acceleration in the source. Therefore, the detection of high-energy neutrinos accompanying photons represents direct evidence for the hadronic model of blazars.

The multiwavelength association of blazars and neutrinos is greatly facilitated by the fact that their emission is highly variable on different timescales, from flares that last minutes to days to several months in a high state of radiation. Also, it is easier to identify a point source in a transient search because of the lower background accumulated over the relatively short duration of the burst \cite{pinpoint}. It is noteworthy that AMANDA detected three neutrino events in temporal coincidence with a rare orphan flare of blazar 1ES 1959+650 \cite{Resconi and Auckerman}. No attempt was made to evaluate a posteriori statistics for this event although its significance by any account exceeds that of the coincidences presently under discussion.

In June 2015, blazar 3C 279 underwent an intense flare observed by Fermi LAT. The gamma-ray flux increased up to forty times over the steady flux and developed a relatively hard spectrum during the flare \cite{ATel Fermi}. An in-depth study of the data revealed an even higher photon flux and a harder spectral index \cite{Masaki Fermi Symp, TheFermi-LAT:2016dss}. An increase in X-ray emission was observed by SWIFT \cite{swift flare}. The event represents an extraordinary opportunity to investigate the pionic origin of the gamma rays by identifying temporally coincident cosmic neutrinos in IceCube. IceCube data is routinely subjected to a blind analysis. The data covering this event, and others discussed in this paper, have not been unblinded, which follows a yearly process.  

 In this paper, we investigate the prospects of observing muon neutrinos in coincidence with this and other flares of blazar 3C 279 by calculating the number of neutrino events based on estimates previously developed in connection with the 2002 burst of blazar 1ES 1959+650. We also comment on a recent flare of this blazar. Note that we focus on muon neutrinos, whose directions can be reconstructed with a resolution of $0.3^{\circ}$, allowing for statistically compelling coincidences that are unlikely to emerge with electron and tau neutrinos, which at present are only reconstructed to about $10^{\circ}$.
\section{Neutrino flux from a pionic gamma-ray source}
In the hadronic scenario, MeV-TeV gamma rays are produced from protons colliding with radiation or gas surrounding the object. These collisions generate charged and neutral pions which decay, producing high-energy gamma rays and neutrinos. Here, we follow estimates \cite{Halzen&Hooper} that relate the neutrino flux to the observed gamma-ray flux using energy conservation:
\begin{equation}
\int^{E_{\gamma}^{\rm{max}}}_{E_{\gamma}^{\rm{min}}} E_{\gamma} \frac{dN_{\gamma}}{dE_{\gamma}} dE_{\gamma} = K \int^{E_{\nu}^{\rm{max}}}_{E_{\nu}^{\rm{min}}} E_{\nu} \frac{dN_{\nu}}{dE_{\nu}} dE_{\nu},
\label{compare}
\end{equation}
 where the factor $K=1(4)$ for $pp(p\gamma)$ interactions. Considering multipion interaction channels, $K$ is changed to approximately 2 in the case of $p\gamma$ interactions.

 The proton spectrum, resulting from Fermi acceleration, as well as the accompanying photon and neutrino spectra are expected to follow a power law spectrum with $\alpha \approx 2$. However, the observed photon spectrum steepens because the gamma rays are absorbed by propagation in the EBL and, possibly, also in the source. Therefore, the observed gamma-ray spectrum is assumed to follow
\begin{equation}
\frac{dN_{\gamma}}{dE_{\gamma}} = A_{\gamma} E_{\gamma}^{- \alpha},
\label{gammaspec}
\end{equation}
 with $\alpha > 2$. In contrast to the photons, the neutrino spectrum is not modified by absorption.
 Although one may propagate the observed gamma rays in the EBL to find the de-absorbed spectrum, it is in general not possible to match the neutrino spectrum to the gamma rays because of gamma rays cascading inside the source. However, energy is conserved in the process, and the total energy between neutrinos and photons can still be related. Even this will result in a lower limit on the neutrino flux because photons absorbed in the source are not accounted for in the left hand side of Eq. \ref{compare}. Using Eq. \ref{compare} and assuming $E_{\gamma, \rm{max}} \gg E_{\gamma, \rm{min}}$, we obtain the following neutrino spectrum:
\begin{equation}
\frac{dN_{\nu}}{dE_{\nu}}   \approx  A_{\nu}  \, E_{\nu}^{-2} \approx \frac{A_{\gamma} E_{\gamma, \rm{min}}^{-\alpha+2}}{(\alpha-2) K \ln{(E_{\nu,\rm{max}}/E_{\nu,\rm{min}})}}  \, E_{\nu}^{-2}.
\label{nuflux}
\end{equation}
 Here, $E_{\gamma, \rm{min}}$ is the minimum energy of photons reflecting the threshold energy of pion production in $pp$ interactions and for the production of the delta resonance in $p\gamma$ interactions. For $pp$ collisions, the minimum energy required for pion production is:
\begin{equation}
E_{p}^{\rm{min}} = \Gamma \, \frac{(2 m_p + m_{\pi})^2 -2 m_p^2 }{2 m_p} \simeq \Gamma \times  1.23 \, \rm{GeV},
\label{Emin-pp}
\end{equation}
where $\Gamma$ is the Lorentz factor of the jet relative to the observer. Given that three pions are produced and that, on average, each charged pion produces four leptons and each neutral pion two photons, the relation between proton energy and gamma-ray and neutrino energies is:
\begin{equation}
E_{\gamma}^{\rm{min}}=\frac{E_p^{\rm{min}}}{6}, \,\,\, E_{\nu}^{\rm{min}}=\frac{E_p^{\rm{min}}}{12}.
\end{equation}
 For $p\gamma$ collisions, the energy threshold is set by the delta resonance $p \gamma \rightarrow \Delta \rightarrow \pi N$:
\begin{equation}
E_{p}^{\rm{min}} = \Gamma^2 \, \frac{m_{\Delta}^2 - m_p^2}{4 E_{\gamma}} \simeq \Gamma^2 \, \bigg(\frac{1 \,\rm{MeV}}{E_{\gamma}}\bigg)  \times 160 \,\rm{GeV}.
\label{Emin-pgamma}
\end{equation}
 In this case, the gamma-ray and neutrino energies are related to the proton energy by
\begin{equation}
E_{\gamma}^{\rm{min}}=\frac{E_p^{\rm{min}} <x_{p \rightarrow \pi}> }{2}, \,\,\, E_{\nu}^{\rm{min}}=\frac{E_p^{\rm{min}} <x_{p \rightarrow \pi}>}{4},
\end{equation}
where $<x_{p \rightarrow \pi}> \simeq 0.2$ is the average fraction of proton's energy transferred to the pion.
\section{Neutrinos in coincidence with 3C 279 Flares}
Blazar 3C 279 is a flat spectrum radio quasar (FSRQ) located at declination $-5.8 ^{\circ}$ and right ascension $194^{\circ}$ with a redshift of 0.536. It is one of the brightest sources in the EGRET catalogue \cite{EGRET} and was the first FSRQ discovered at TeV energy by MAGIC in 2006 \cite{MAGIC}.  The MAGIC collaboration has reported a gamma-ray flux from 3C 279 of $5.2 \times 10^{-10} \, TeV^{-1}cm^{-2}s^{-1}$ with a spectral index of 4.1. The EBL corrected spectrum has a smaller spectral index of 2.94.

 Blazar 3C 279 has consistently exhibited rapid variations in flux, and multiple flares have been observed. On June 16, 2015, Fermi observed an intense flare of GeV gamma rays from 3C 279 reaching forty times the steady flux of this source. The spectral study of this flux found a relatively hard spectral index. Specifically, we will use in our estimates the average daily photon flux of $24.3\times 10^{-6}\, ph\,cm^{-2}\,s^{-1}$ with a spectral index of 2.1.

\begin{figure}
\centering\leavevmode
\includegraphics[width=4.0in]{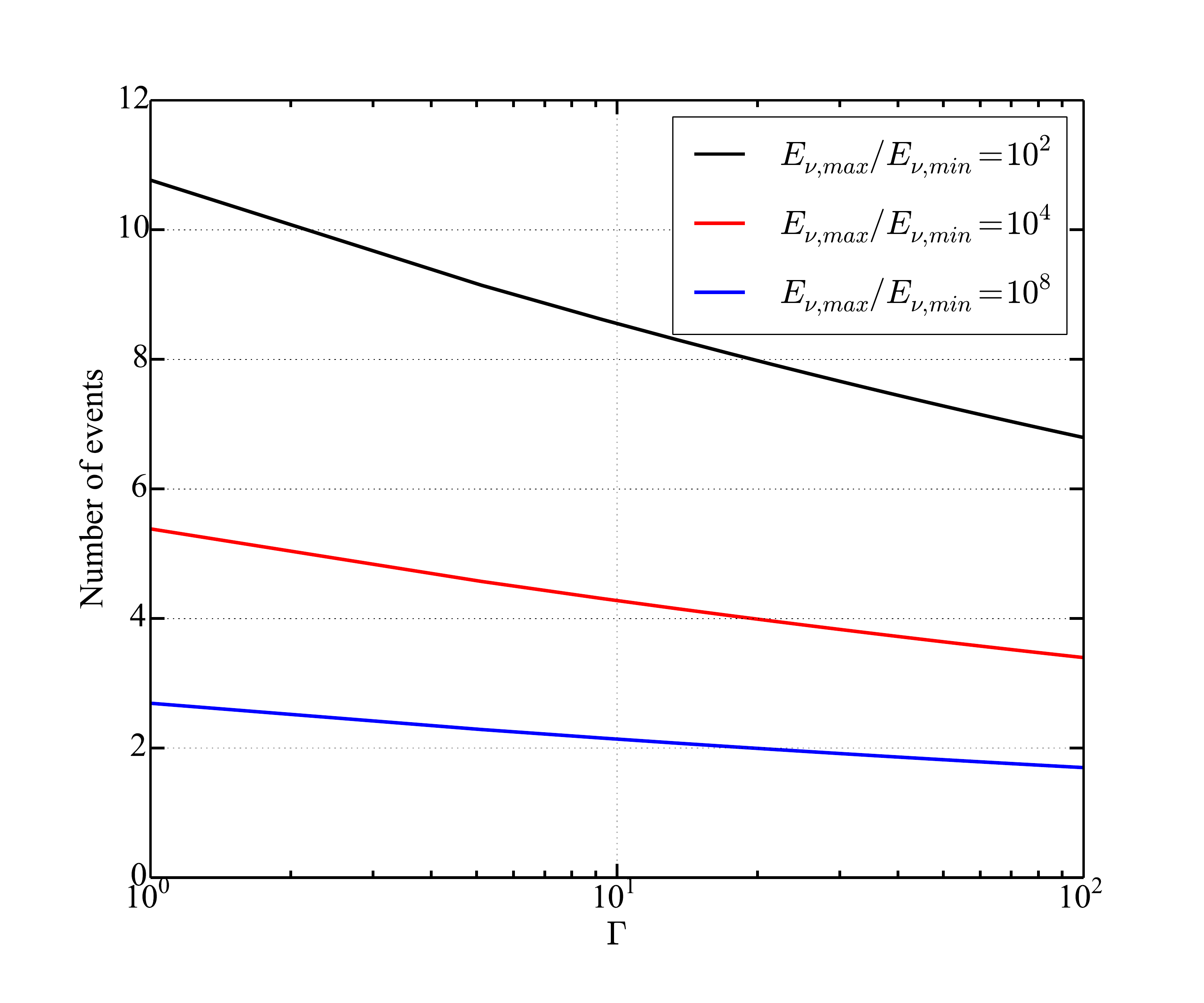}
\caption{Estimated number of events from blazar 3C 279 flare in June 2015 for different energy ranges of neutrino emission from $pp$ collision. The events correspond to neutrino energies above 1 TeV.}
\label{pp-2d}
\end{figure}

\begin{figure}
\centering\leavevmode
\includegraphics[width=4.0in]{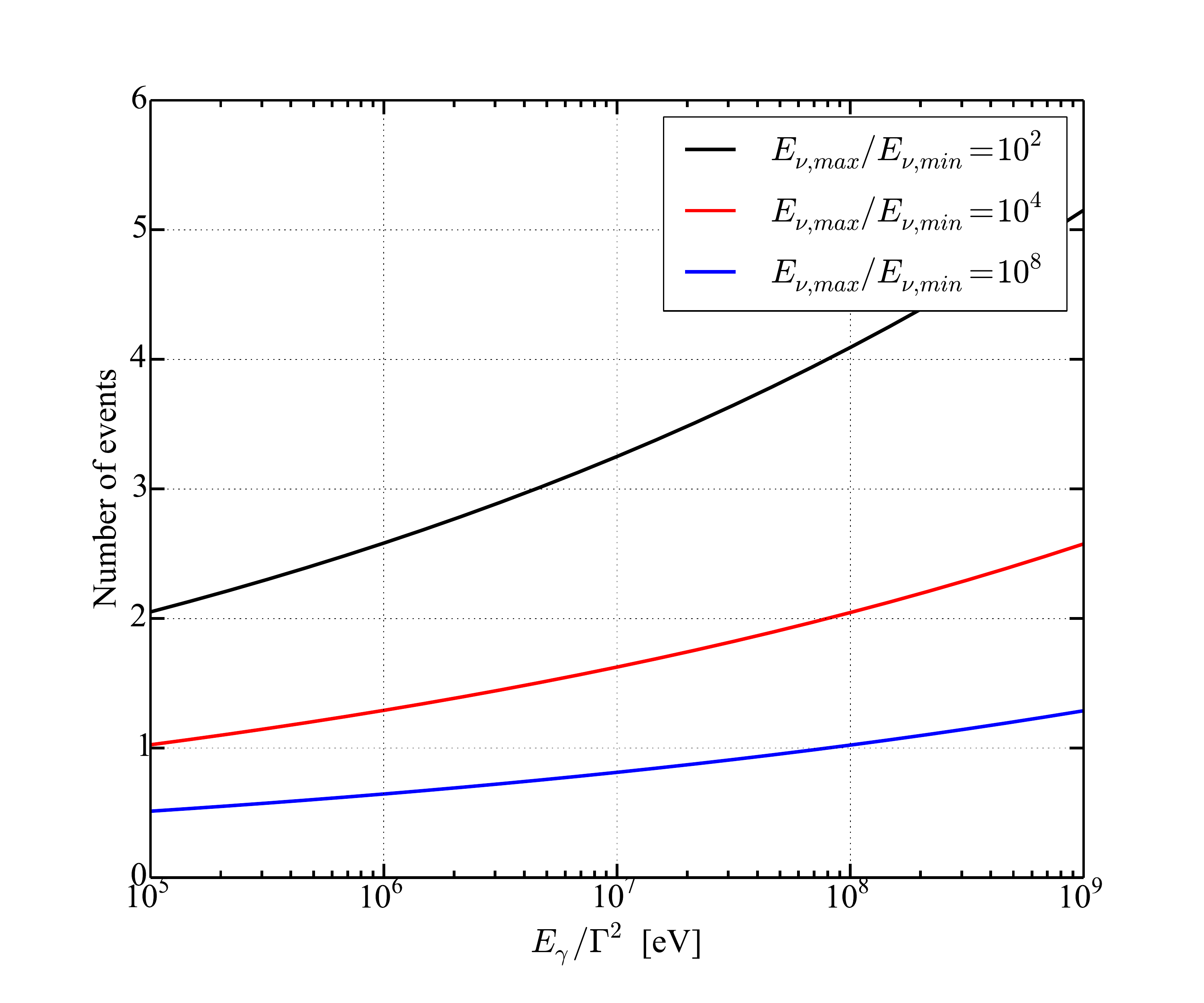}
\caption{Estimated number of events from blazar 3C 279 flare in June 2015 for different energy ranges of neutrino emission from $p\gamma$ collision. The events correspond to neutrino energies above 1 TeV.}
\label{pgamma-2d}
\end{figure}

\begin{figure}
\centering\leavevmode
\includegraphics[width=4.0in]{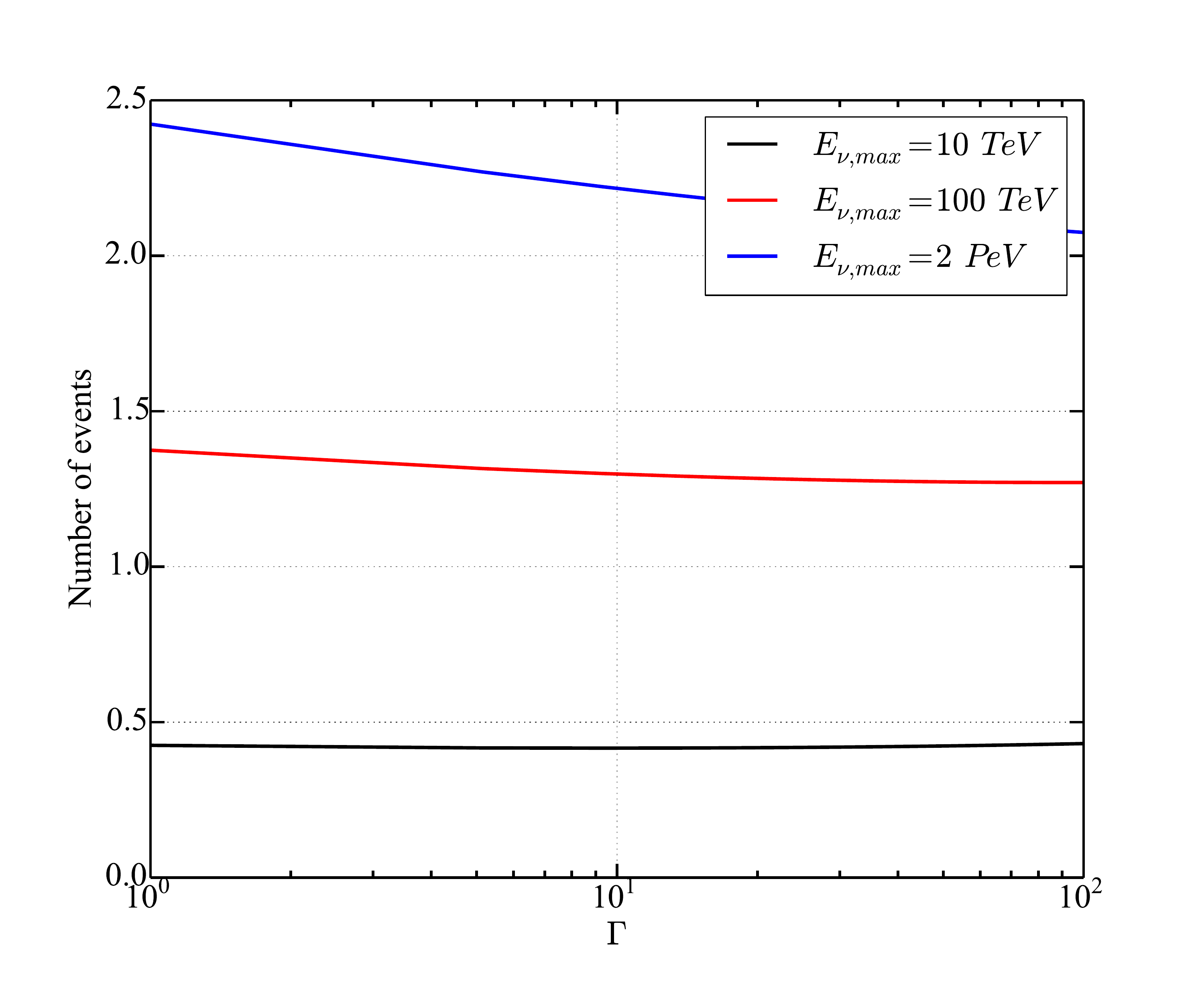}
\caption{Estimated number of events from blazar 3C 279 flare in June 2015 for different values of maximum neutrino energy in $pp$ collision when neutrinos and gamma rays are produced in the same energy range. The events correspond to neutrino energies above 1 TeV.}

\label{pp-2d-2}
\end{figure}

\begin{figure}
\centering\leavevmode
\includegraphics[width=4.0in]{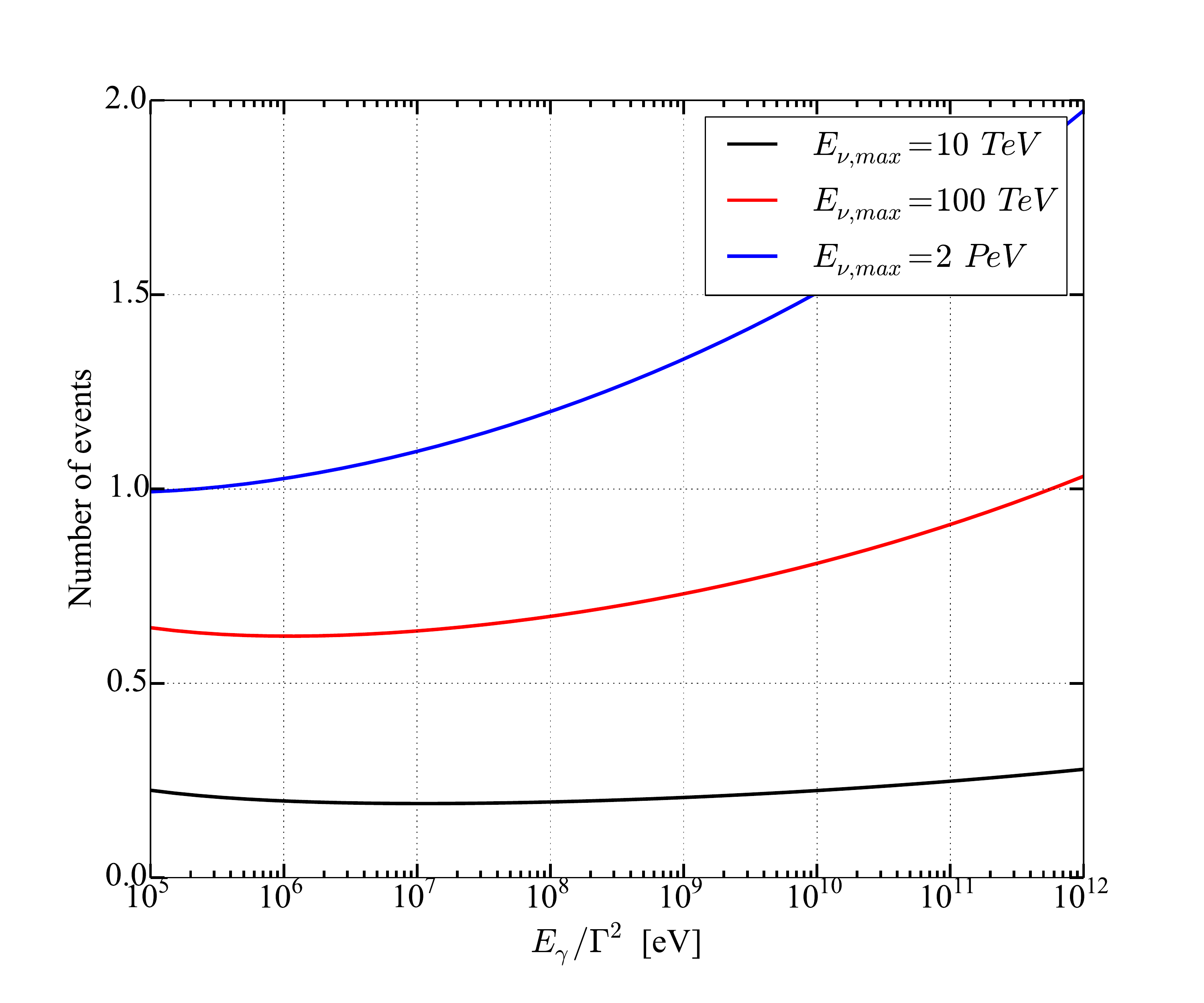}
\caption{Estimated number of events from blazar 3C 279 flare in June 2015 for different values of maximum neutrino energy in $p\gamma$ collision when neutrinos and gamma rays are produced in the same energy range. The events correspond to neutrino energies above 1 TeV.}
\label{pgamma-2d-2}
\end{figure}

\begin{figure}
\centering\leavevmode
\includegraphics[width=4.0in]{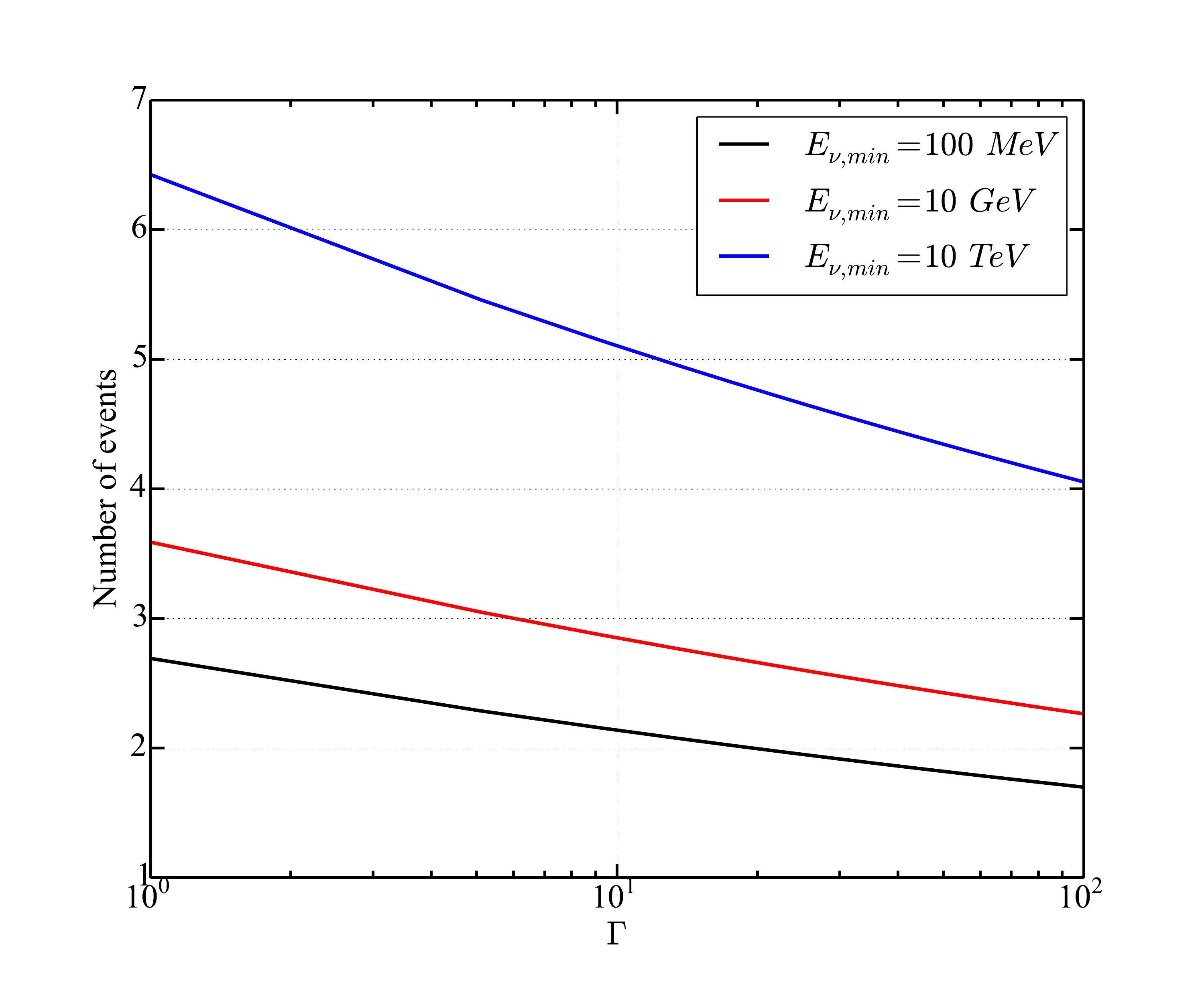}
\caption{Estimated number of events from blazar 3C 279 flare in June 2015 for different minimum energy ranges of neutrinos in $pp$ collision, assuming that the maximum neutrino energy is 10 PeV. The events correspond to neutrino energies above 1 TeV.}
\label{pp-2d-3}
\end{figure}

\begin{figure}
\centering\leavevmode
\includegraphics[width=4.0in]{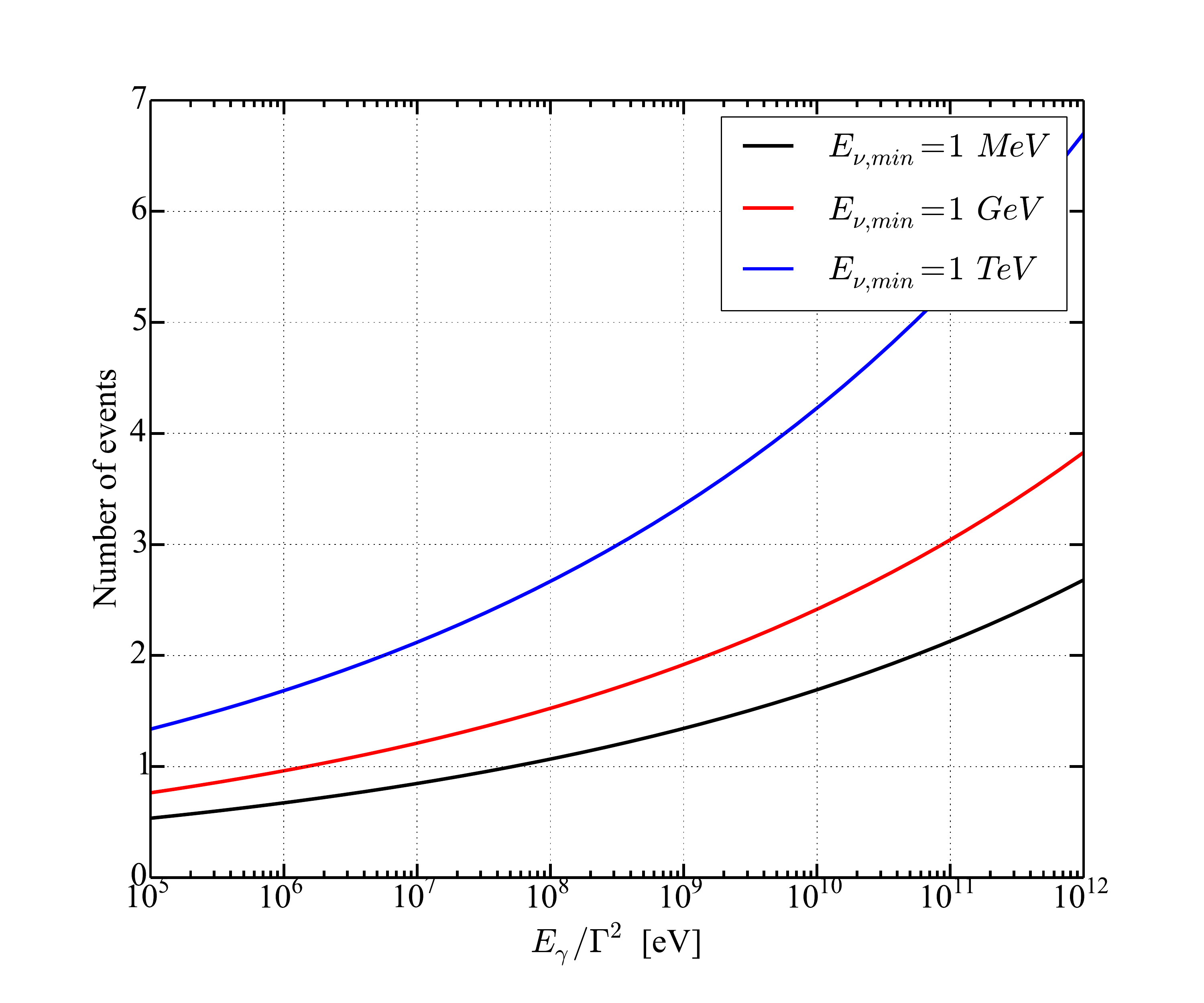}
\caption{Estimated number of events from blazar 3C 279 flare in June 2015 for different minimum energy ranges of neutrinos in $p\gamma$ collision, assuming that the maximum neutrino energy is 10 PeV. The events correspond to neutrino energies above 1 TeV.}
\label{pgamma-2d-3}
\end{figure}

 The expected number of well-reconstructed muon neutrinos in IceCube is calculated using Eq. \ref{nuflux} and 
\begin{eqnarray}
N_{\nu_\mu+\overline{\nu}_\mu} = t \int \frac{dN_\nu}{dE_\nu}\,A_{eff}(E,\theta)\,dE
\label{Nevent}
\end{eqnarray}
where the effective area $A_{eff}$ is taken from \cite{IC 4yr PS study}.
The neutrino flux calculated from the energy balance relation of Eq. \ref{compare} depends on the value of $E_{\nu,\rm{max}}/E_{\nu,\rm{min}}$, which represents the energy interval over which proton interactions produce pionic gamma rays. We will consider three possible values for this ratio:
\begin{itemize}
\item Case 1: 
We assume that neutrinos are exclusively produced in a specific energy range. This is similar to the approach in reference \cite{TANAMI} where the neutrino spectrum is assumed to peak at PeV energies. The number of events for various values of $E_{\nu,\rm{max}}/E_{\nu,\rm{min}}$ is shown in Fig. \ref{pp-2d}(\ref{pgamma-2d}) for $pp(p\gamma)$ interactions. For $pp$ collisions, the number of neutrinos expected is within IceCube's sensitivity for the wide range of values for the Lorentz factor and the neutrino energy range considered. For $p\gamma$ collisions, large values of $E_\gamma/\Gamma^2$ are required for observing the flare in neutrinos. Note that for $p\gamma$ interactions we label the energy in terms of the ratio $E_\gamma/\Gamma^2$. The actual energy depends on the value of $\Gamma$ which is often not directly measured and must be obtained from further modeling of the spectrum \cite{TheFermi-LAT:2016dss,Paliya:2015ufo}.
\item Case 2: We assume that neutrinos are produced over the same energy range as the gamma rays, i.e., $E_{\nu, min}$ is obtained from Eq. \ref{Emin-pp}(\ref{Emin-pgamma}) for $pp(p\gamma)$ collisions. The resulting number of neutrino events is shown in Figs. \ref{pp-2d-2} and \ref{pgamma-2d-2} for different values of maximum neutrino energy. Here, the estimated number of events strongly depends on the maximum energy achieved by the cosmic accelerator.
\item Case 3: Finally, we consider different threshold energies for neutrinos, assuming that the maximum neutrino energy is 10 PeV, the highest energy observed by IceCube so far. The results are shown in Fig. \ref{pp-2d-3}(\ref{pgamma-2d-3}) for $pp$($p\gamma$) collisions. Notice that a higher minimum neutrino energy corresponds to a larger number of events observed.
\end{itemize}
%
%
\section{Discussion and Conclusion}
The above calculations illustrate that blazar flares represent an extraordinary opportunity to identify the origin of IceCube neutrinos. The short time window results in a lower number of events but also a suppressed background. For the specific burst of 3C 279, we have shown that there is a clear opportunity for observing coincident neutrinos, especially in the case of $pp$ interaction. 

 In addition to the intense flare in June 2015, two previous flares were observed during December 2013 and April 2014 \cite{Masaki2015}. Although their photon flux is not as large as for the flare discussed above, stacking them will result in a higher likelihood of finding neutrinos. The average daily photon flux of all three flares is listed in Table \ref{flux_table_day}. Assuming, for simplicity, that the neutrino spectrum would follow the gamma-ray spectrum, the total number of events for a flat spectrum of neutrinos will be about 4(2) for $pp$($p\gamma$) collisions.
\begin{table}[h]
    \begin{center}
\begin{tabular}{|c|c|c|c|}
  \hline
    Date & $F_\gamma \, [ph \, cm^{-2}s^{-1}]$  \\
    \hline
    December 20, 2013 & $6\times10^{-6}$  \\
    \hline
    April 3, 2014 & $6.4\times10^{-6}$  \\
    \hline
    June 16, 2015 & $24.3\times10^{-6}$  \\
 \hline
\end{tabular}
\end{center}
    \caption{The date and average daily photon flux of observed flares from 3C 279. All fluxes are measured above 100 MeV \cite{Masaki Fermi Symp,TheFermi-LAT:2016dss}. }\label{flux_table_day}
    \end{table}

We have also estimated the number of events for each flare using detailed information of fluxes and assuming same spectral behavior for neutrinos and gamma rays. Detailed duration and flux are listed in Table \ref{flux_table}. The total number of events obtained is 4(2) for $pp$($p\gamma$) collisions. Statistics are straightforward with less than 0.001 background of atmospheric events per day within the resolution of $0.3^\circ$. 
\begin{table}[h]
    \begin{center}
\begin{tabular}{|c|c|c|c|c|c|c|}
  \hline
    Date & $F_\gamma \, [ph \, cm^{-2}s^{-1}]$ & $\alpha$ & Duration [day]   \\
    \hline
    December 20, 2013 & $11.71\times10^{-6}$ & 1.71 & 0.2  \\
    \hline
    April 3, 2014 & $11.79\times10^{-6}$ & 2.16 & 0.267   \\
    \hline
    June 16, 2015 & $24.3\times10^{-6}$ & 2.1 & 1  \\
 \hline
\end{tabular}
\end{center}
    \caption{The date, photon flux, spectral index, and duration of observed flares from 3C 279. All fluxes are measured above 100 MeV \cite{Masaki2015}. }\label{flux_table}
    \end{table}

 The FSRQ 3C 279 is included in the IceCube source list for both time-dependent \cite{time dep paper} and time-independent point source searches \cite{IC 4yr PS study}. The latest time-dependent search looked for a correlation of neutrinos with observed flares up until 2012. This period did not include any flares from 3C 279. Future time-dependent studies may reveal signals from these flares.

 Recalling the temporal coincidence observed in AMANDA with flares of blazar 1ES 1959+650, it is noteworthy that a new very high energy flare from 1ES 1959+650 was observed by VERITAS during October 2015 \cite{ATEL_VERITAS}. According to the preliminary analysis of the data, the flux has reached $\sim$ $50\%$ of the Crab flux with a spectral index of 2.5. Detailed analysis will provide more information about the duration and spectrum of this flare. Based on the preliminary results, and provided that the gamma rays are hadronic in origin, IceCube expects to observe $\sim$ 0.1 events per hour for this burst. If the burst has lasted for more than a day, then it is very likely that accompanying neutrinos would be observed in IceCube. Its location is obscured by the earth, and the highest energy events will therefore be absorbed. 
\section{Acknowledgments}
We would like to thank Markus Ahlers, Keith Bechtol, Markus Ackermann, and Marcos Santander for useful discussions and their informative comments during this study. This research was supported in part by the U.S. National Science Foundation under Grants No. ANT-0937462 and PHY- 1306958 and by the University of Wisconsin Research Committee with funds granted by the Wisconsin Alumni Research Foundation.

\end{document}